\documentclass[%
  twocolumn,
 amsmath,amssymb,
 aps,
 prl,
]{revtex4-1}

\usepackage{graphicx}
\usepackage{dcolumn}
\usepackage{bm}
\usepackage{hyperref}
\usepackage{amsmath}
\usepackage{amssymb}
\newcommand{\mb}[1]{ \mbox{\boldmath$#1$} }

\newcommand{\ds}{\displaystyle}
\newcommand{\beq}{\begin{eqnarray}}
\newcommand{\eeq}{\end{eqnarray}}
\newcommand{\beqq}{\begin{eqnarray*}}
\newcommand{\eeqq}{\end{eqnarray*}}
\newcommand{\p}{\partial}

\newcommand{\eps}{\varepsilon}
\newcommand{\x}{\mbox{\boldmath$x$}}

\newcommand{\y}{\mbox{\boldmath$y$}}

\newcommand{\n}{\mbox{\boldmath$n$}}

\font\bb=msbm10 at 12pt
\def\rR{\hbox{\bb R}}

\begin{document}
%%%%%%%%%%%%%%%%%%%%%%%%%%%%%%%%%%%%%%%%%%%%%%%%%%%%%%%%
\title{Triangulation sensing: how cells recover a source from diffusing particles in three dimensions}% Force line breaks with \\
%%%%%%%%%%%%%%%%%%%%%%%%%%%%%%%%%%%%%%%%%%%%%%%%%%%%%%%%
\author{Ulrich Dobramysl$^{1}$  and David Holcman$^{2}$}
\affiliation{$^{1}$ Wellcome Trust/Cancer Research UK Gurdon Institute, University of Cambridge, Tennis Court Rd, Cambridge CB2 1QN,United Kingdom $^{2}$ Group of data modeling and computational biology, PSL, Ecole Normale Superieure, Paris, France}

\begin{abstract}
How can cells embedded into a gradient concentration triangulate the position of the source and migrate toward their final destination?
The source triangulation requires to recover the three dimensional coordinates of the source from the fluxes of diffusing cues at narrow windows (receptors) located on the surface of a cell. We develop here a method to address this question and we show in the limit of fast binding rate to the receptors, that at least three receptors are necessary. We solve the steady-state diffusion equation using an asymptotic approach, which agrees with hybrid stochastic-analytical simulations. Interestingly, with an accuracy of few percent, the source cannot be located if it is located at a distance tens of time the size of the cell.  Finally, the precision of the source recovery increases with the number of receptors.
\end{abstract}
%\keywords{Suggested keywords}%Use showkeys class option if keyword
                           %display desired
\maketitle
%\tableofcontents
How can a cell find a target position with a sufficient precision? This question is ubiquitous biology: Spermatozoa need to find the egg in the uterus \cite{kaupp2017signaling,strunker2015physical,wachten2017sperm,alvarez2014computational}; during brain wiring, axon migrates to their final location interpreting, with a yet unknown mechanism, a complex ensemble of molecular gradients (Fig. \ref{fig:figure1}A) \cite{chedotal2010wiring,goodhill2016can,reingruber2014computational};  Combining chemical and mechanical gradients cell need to migrate to a small target in complex environments \cite{gorelashvili2014amoeboid,wondergem2019chemotaxis}.\\
The physical principles by which a cell could find the location of a target remains an open question. The first step is the ability of the cell to detect a difference of gradient concentration across its few micron in size. This was the subject of Berg and Purcell approaches \cite{Berg1977,Zwanzig,bialek2008cooperativity,endres2008accuracy,kaizu2014berg,aquino2016know}, based on the computation of the difference in concentration across the surface of a small diffusing ball. However, this paradigm is not sufficient to decipher how a cell is able to triangulate the exact position of a gradient source, which goes beyond detecting a gradient concentration \cite{dobramysl2018mixed,dobramysl2018reconstructing,shukron2019chemical}. Here, the physical model consists instead of many receptors distributed across the surface of a cell that bind diffusing molecules at a fast rate. The flux imbalances between different receptors then form a directional signal from which the cell could triangulate the position of the source thereby identifying its exact location. Triangulating a source in two dimensions required at least three small receptors to reconstruct the location of the source \cite{dobramysl2018reconstructing}. However this reconstruction is possible only if the source is exceedingly close -- only up to 20 to 40 times the cell diameter. In that case, the level of detection sensitivity decreases with the reciprocal of the distance to the source. Yet, the growthcone, which is the tip of an axon has a size of few microns, but nevertheless is able to accurately find its final destination over rather long distances (mm to cm). This paradox was resolve by understanding that migration is constraint in narrow tubes, where triangulation works at much longer ranges due to the asymmetric location of the source \cite{dobramysl2018reconstructing}, which effectively reduces the search to one dimension. Finally, another interesting property of triangulation is the consequence of many redundant receptors which results in increasing the accuracy of localisation and can thus compensate for possible fluctuation in the receiving fluxes.\\
In the present letter, we study the triangulation sensing in three dimensions using a diffusion model for the cues. We provide an analytical expression for the receptor fluxes using the method of matched asymptotics for the Laplace operator. Using this solution, we determine how the location of the source depends on the position of the receptors. We numerically determine the position of the source for various receptor placement configurations and show that this can lead to several order of magnitude difference in the sensitivity and susceptibility to noise. Surprisingly, contrary to the two dimensional case, the sensitivity decays with distance square between the source and the target. Finally, having many redundant receptors on the cell surface drastically reduces the possible fluctuations in the fluxes.\\
{\bf \noindent Diffusion model of cell triangulation.} The model (Fig. \ref{fig:figure1}B) consists of diffusing cues that have to bind to $M$ narrow windows located on the surface of three dimensional ball $B_a$ of radius $a$. Individual cue molecules are described as Brownian particles. The cues are released from a source at position $\x_0$ outside the ball (Fig.~\ref{fig:figure2}A). Our goal is to estimate the steady-state flux at each narrow window for fast binding (i.e. the probability density has an absorbing boundary condition at the windows). The first step is to solve the Laplace equation
\beq
D\Delta P_0(\x) & = & -\delta_{\x_0} \hbox{ for } \x\, \in \,\rR^3 -B_a \label{eqDP3b}
\eeq
where $\frac{\p P_0}{\p \n}(\x)=0$ for $\x\in\p B_a-S_k(\eps)$, where $S_k(\epsilon)$ are nonoverlapping circular windows representing receptors (or possibly clusters thereof) of radius $\eps$ centered at points $x_k$ on the surface of the sphere; the remaining surface of the sphere is absorbing with $P_0(\x)=0$ for $\x\in\Sigma_a=S_1(\eps)\cup...\cup S_n(\eps)$. As $\x$ tends to infinity, the gradient needs to dissipate, hence we have the additional condition $\lim_{|\x|\rightarrow\infty}P_0(\x)=0$. Using the Green's function exterior to a ball in three dimensions, a solution of \ref{eqDP3b} can be found using Neumann's function~\cite{lagache2017extended,lindsay2017first},
{\small
\beq \label{Neumann}
\mathcal{{N}}(\x,\x_0)&=& \ds\frac{1}{4\pi |\x-\x_0|}+\frac{a}{4\pi |\x_0||x-\ds \frac{a^2 \x_0}{|\x_0|^2}|}\nonumber \\&+&
\ds{\frac{1}{4\pi a }\log\left( \ds\frac{\ds\frac{|\x_0||\x|}{a^2}\left(1-\cos(\theta)\right)}{\tilde d(\x,\x_0)}\right)},
\eeq}
where {\small $\tilde d(\x,\x_0)= 1-\frac{\x_0.\x}{a^2} +\left(1+(\frac{|\x_0||\x|}{a^2})^2-2\frac{\x_0.\x}{a^2} \right)^{\frac{1}{2}}$} and $\theta= \sphericalangle \x_0 \x$.
$N$ is solution of the Laplace's equation
\beq
\Delta \mathcal{{N}}(\x,\x_0)&=&-\delta(\x-\x_0) \hbox{ for } \x \in \mathbb{R}^3 \nonumber \\
\frac{\p \mathcal{{N}}}{\p n} (\x,\x_0) &=& 0 \hbox{ for } \x \in S_a=\p B_a.
\eeq
where $S_a$ is the surface of the three-dimensional ball $B_a$ and $\x_0 \in \rR^3-B_a$ the location of the source. Finally, 
\beq \label{eqgen}
P_0(\x)= N(\x,\x_0)+\sum_{k} \int_{S_k(\eps)} \mathcal{N}(\x,\x_k) \frac{\p w(\x)}{\p n} dS_{\x}.
\eeq
Defining the vector $\tilde{\mb{\alpha}}$ with the entries
$\alpha_i=-\mathcal{N}(\x_i,\x_0)$ and the unknown fluxes on each window
\beq
\frac{\p P_0}{\p \n}(\y) =\frac{A_k}{\sqrt{\eps^2-r^2}}, \hbox{ for } \y \in S_k(\eps),
\eeq
using eq. \ref{eqgen}, the fluxes satisfy the following matrix equation:
%%%%%%%%%%%%%%%%%%%%%%%%%%%%%%%%%%%
\begin{figure}
  \centering
  \fbox{\includegraphics[width=\columnwidth]{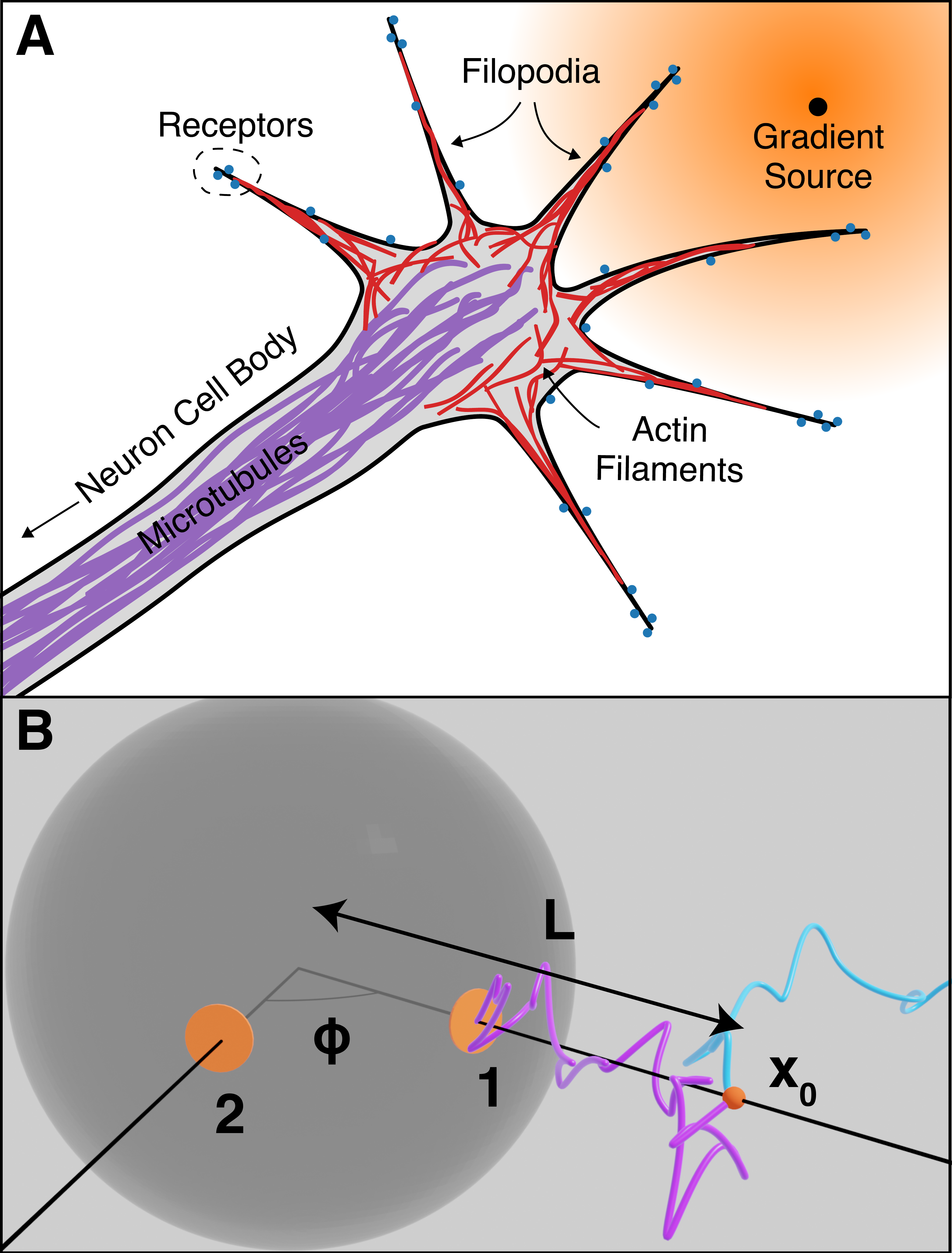}}
  \caption{ (A)Drawing of a neuronal growth cone in an external chemical
    gradient. Receptors on the cell membrane are able to sense the external chemical
    gradient, thereby defining navigation in the Brain. (B) model of the sensing part of a cell as a ball containing narrow windows (receptors). The source is located at position $\x_0$ where Brownian particles are released (blue and purple trajectories).}
  \label{fig:figure1}
\end{figure}
%%%%%%%%%%%%%%%%%%%%%%%%%%%%%%%%%%%
\beq \label{sysMatrix2}
     [\mb{\tilde M}]{\mb{ \tilde A}}=\tilde{\mb{\alpha}}
\eeq
where $[\mb{\tilde M}]=\theta_\eps\mb{I}+2\eps/\pi\mb{N}$ and
$\theta_\eps=\pi/2+\eps\log(\eps/a)/(2a)+B\eps$, where $B$ is a constant (third order in the asymptotic expansion) will be determined numerically depending on the window locations. The symmetric matrix $\mb{N}$ has zeros on its diagonal and the remaining entries are given
by $[\mb{N}]_{ij}=\mathcal{N}(\x_i,\x_j)$ where $i,j=1...M$, $i\neq
j$. $\mathcal{N}(\x,\y)$ is the Neumann-Green's function defined by \ref{Neumann}. The structure of the matrix $\mb{\tilde M}$ disallows an explicit solution for any number of windows $m$, but can be inverted for a low number of window $m$. Finally, after numerically solving equation~\ref{sysMatrix2}, the flux on each window is $\Phi_k=2\pi A_k$. To first order, this can be solved analytically as
\beq
  \label{flux}
  \Phi_k=\theta_{\eps}^{-1}\!\! \left(\!\! \alpha_k \!-\! \frac{2 \pi \eps}{\theta_\eps}\sum_{q\neq k}\mathcal{N}(\x_q,\x_k)\alpha_k \!\!\right)\!\! + \!O\!\left(\left[\frac{2 \pi \eps}{\theta_\eps}\right]^2\right)\!\!.\quad
\eeq
We first confirm the validity of this result using a steady-state hybrid stochastic simulation scheme to compute these fluxes: briefly, after Brownian particles are released from the source $S(\x_0)$, their position is immediately mapped to the surface of a sphere $\p B_{R}$ around the ball $B_a$ as $R\gg a+\eps$ (Fig. \ref{fig:figure2}A). The probability distribution of this mapping is given by
\beq
P(\x;\x_0)=\frac{1}{4\pi}\frac{\beta^2-1}{(1+\beta^2-2\beta\cos\gamma)^{3/2}},
  \label{eq:mapping}
\eeq
which represents the flux through the absorbing boundary $\p
B_{R}$ in free space, here $|\x|=R$, $|\x||\x_0|\cos\gamma=\x\cdot\x_0$ and $\beta=|\x_0|/R$. After mapping the position, a particle performs a standard Brownian motion (Euler-Maruyama scheme) until it is absorbed by a window \cite{dobramysl2018mixed}. A particle can also leave the test ball of radius $R_e>R$ (this is to prevent frequent mappings), upon which it is mapped back to the surface $\p B_{R}$ using Eq.~(\ref{eq:mapping}). For each mapping, a particle has a finite chance $P_e=R/|\x_0|$ to escape to infinity, whereupon its trajectory is terminated.\\
%%%%%%%%%%%%%%%%%%%%%%%%%%%%%%%%%%%%%%%
\begin{figure}
    \centering
    %\fbox{\rule{10cm}{0pt}\rule[-0.5ex]{0pt}{8cm}} %placeholder
    \includegraphics[scale=0.8]{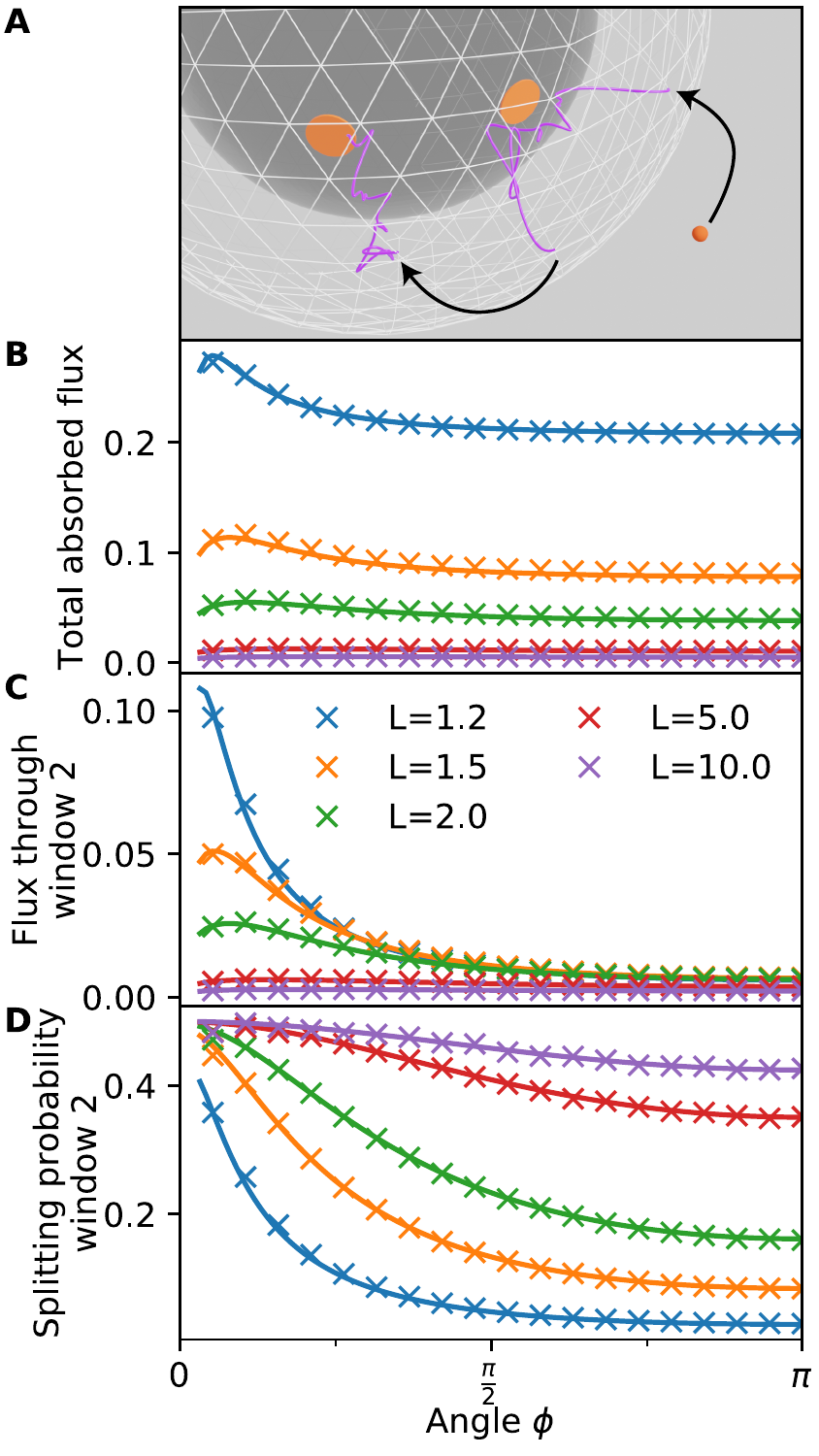}
    \caption{(A) Schematic of a reflecting ball with window one
      (orange disk) facing the source (small orange sphere) and a second window
      at an angle $\phi$. Particles are released by the source undergo
      Brownian motion (hybrid simulation) and are either absorbed by one of the windows
      (trajectory in magenta) or can escape to infinity. (B) Total flux through both windows vs the angle $\phi$ , (C) absolute flux through window two and (D) splitting probability for a particle to hit window two. Curves are for various distance $L$ to the source: analytical results eq. \ref{flux} (solid lines), compared to simulation data (crosses).}
    \label{fig:figure2}
\end{figure}
%%%%%%%%%%%%%%%%%%%%%%%%%%%%%%%%%%%%%%%
{\bf \noindent Accuracy in recovering only the direction of a source.}
To illustrate the model, we start with two windows located on the equator of the ball (Fig.~\ref{fig:figure2}A): window one faces the source directly and window two is at an angle $\phi$. The total flux through both windows vs the source distance $L=|\x_0|$ and the angle $\phi$, shows an excellent agreement (Fig.~\ref{fig:figure2}B) between the analytical results (solid lines) and simulated data (crosses), with the single fit parameter $B=5.7$. The flux through window two alone
(Fig.~\ref{fig:figure2}C) shows a single maximum when the windows are
close and decays rapidly to a small but finite value when the window
is on the opposite side of the ball. The splitting probability $p_S=\frac{\Phi_1}{\Phi_1+\Phi_2}$ of hitting window two given that a particle hits one of the windows allows distinguishing the direction only of the source when it is very close. Already for $L=10$ ball radii, the difference in the hitting probabilities is smaller than $10\%$, which makes any recovery impossible in a noisy environment.\\
%%%%%%%%%%%%%%%%%%%%%%%%%%%%%%%%%%%%%%%
\begin{figure}
    \centering
    %\fbox{\rule{10cm}{0pt}\rule[-0.5ex]{0pt}{8cm}} %placeholder
    \includegraphics{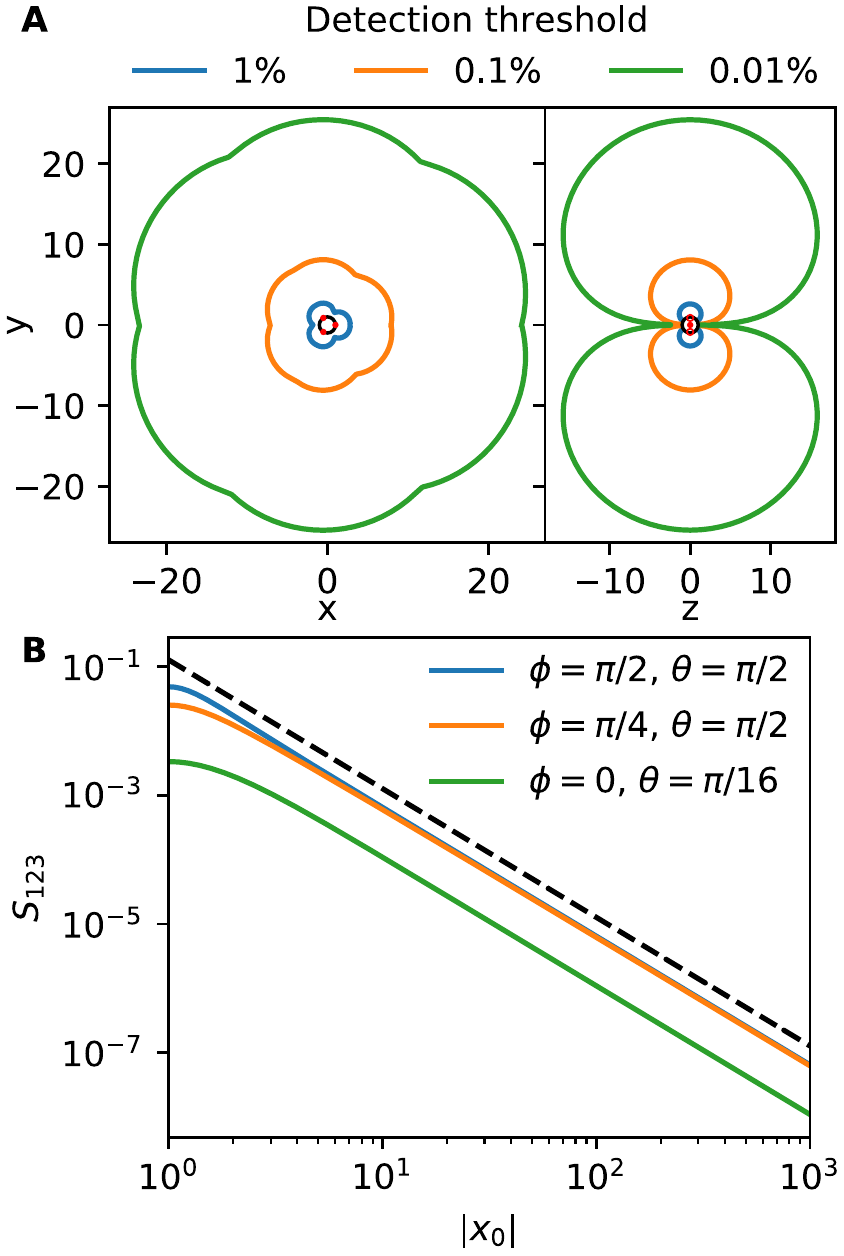}
    \caption{{\bf Sensitivity of detecting the source position from
      Eq.~(\ref{eq:S123})}. (A) for a ball with three windows arranged as an
      equilateral triangle on a geodesic. The detection contours is in
      the plane that contains all 3 windows (Left) and in plane perpendicular to the window plane (Right), for three different detection thresholds (1\%, 0.1\% and 0.01\%). (B) The sensitivity decays with distance of the source $x_0$ like $1/|x_0|^2$.}
    \label{fig:figure3}
\end{figure}
%%%%%%%%%%%%%%%%%%%%%%%%%%%%%%%%%%%%%%%
In order to quantify the distance at which it is still possible to recover the direction of the source with three windows, we use the sensitivity cost function
\beq
\label{eq:S123}
\begin{split}
  f(\x_0;\x_1,\x_2,\x_3)=\max\{&|P_1(\x_0)-P_2(\x_0)|,\\
  &|P_2(\x_0)-P_3(\x_0)|,\\&|P_3(\x_0)-P_1(\x_0)|\},
\end{split}
\eeq
where $\x_0$ is the position of the source and $\x_i$, $i=1, 2, 3$
are the positions of the three windows on $\p B_a$. The cost function $f(\x_0;\x_1,\x_2,\x_3)$  describes the maximum absolute imbalance between the fluxes through the windows. Fig.~\ref{fig:figure3}A shows the contours of this
function for three windows arranged in an equatorial equilateral
triangle in a slice through the $z=0$ and $x=0$ planes at three
different threshold levels. Notably, the distance at which directions
can still be discerned is approximately an order of magnitude less for
any given threshold compared to the equivalent situation in two dimensions ~\cite{dobramysl2018reconstructing}. Indeed, using the dipole expansion for a source located far away$|\x_0|\gg 1$, $f(\x_0;\x_1,\x_2,\x_3)\approx C\frac{\max_{i,j}(|(\x_j-\x_i).\hat{\x_0}|)}{|\x_0|^2}$, where $C>0$ is constant and $\hat{\x_0}=\frac{\x_0}{|\x_0|}$. Fig.~\ref{fig:figure3}B illustrates this decay.\\
{\bf \noindent Triangulating the source location.} 
To reconstruct the location of a source $\x_0$ (three coordinates of the source) from the measured fluxes $\Phi_i$ through the windows located at $\x_i$, we require at least three windows. Interestingly, in two dimensions the minimum number of windows is three as well, even though only two coordinates need to be determined. This is due to the recurrence properties of the Brownian motion, which imposes that the sum of all window fluxes to be unity~\cite{dobramysl2018mixed}. However, this condition is not present in three dimensions because Brownian particles have a finite probability to escape to infinity before hitting a window. The source location $\x_0$ enters Eq.~\ref{sysMatrix2} only via the Neumann-Green's function $\mathcal{N}$, which we invert numerical by rewriting eq.~\ref{sysMatrix2} as a solution of implicit equations
\beqq
F_i(\x_0)=\theta_\eps\Phi_i+\sum_{j\neq i}\mathcal{N}(\x_i,\x_j)\Phi_j-2\pi\mathcal{N}(\x_i,\x_0)=0.
\eeqq
Each of the $m$ equations describes a closed surface in three dimensions, the intersection of which yields the source location. Therefore, we use the following procedure: we search for the joint root of the $F_i(\x_0)$ via first tracing the root contour of $F_1$ in the $x-y$ plane until we find its intersection with the root contour of $F_2$. We then plot the curve described by the joint root contour of
$F_1$ and $F_2$ until $F_3=0$ is fulfilled. This yields the source
location $\x_0$ as a function of the measured fluxes $P_i$ and the
window locations $\x_i$. The choice of window labels used in
this algorithm is arbitrary and we could have used any combination of
three of the $M$ windows .\\
{\noindent \bf Uncertainty of reconstruction in noisy environments}
To study the effects of flux measurement uncertainty on the source position triangulation, we investigate the consequence of small perturbations on the fluxes $\eta\ll\Phi$. The source can be recovered by using the fluxes of any combination of three windows $\Phi_n$, $\Phi_m$ and $\Phi_l$ out of many. We use the Jacobian matrices of the fluxes $J_{ij}=\frac{d\Phi_{i}}{d\x_0^{j}}$ where $i=m$, $n$ or $l$ and invert it to extract the error matrix $\vec{E}_{ij}^{(m,n,l)}=\eta [J_{i j}^{-1}]$. The column vectors of $E^{(m,n,l)}$ contain the linear uncertainty vectors associated with the source location recovery via these windows and the volume of uncertainty for this particular recovery is given by the parallelepiped spanned by these vectors. Because the choice of $m$, $n$ and $l$ is arbitrary, we define the overall volume of uncertainty for a particular configuration of window number, location and source position, as the volume of the geometric intersection of all parallelepipeds resulting from all possible combinations of three windows.\\
We present three cases (Fig. \ref{fig:figure4}): A -- windows spread uniformly across the ball surface, B -- windows concentrated in a single cluster and C -- window clusters spread uniformly across the ball surface. In all cases, adding only a few windows decreases the volume of uncertainty by order of magnitudes, as it correspond to the intersection of an exponential number of parallelepipeds. In all cases, the source located at closest and directly on top of the cell, $\x_0=(0,0,2)$ (blue line), has the least volume of uncertainty. Doubling the distance but staying on top of the ball leads to an increase of roughly three orders of magnitude in the volume of uncertainty in all cases, $\x_0=(0,0,4)$ (yellow line). Overall, case A has high uncertainty for low numbers of windows, however when the number of windows is high, the error depends overwhelmingly on the distance and not on the direction, as expected. In contrast, the single cluster of windows in case B shows a high directional dependence of the error for all windows, with the advantage of low uncertainty even for low numbers of windows. Case C with small clusters spread on the sphere is a compromise between the uniform spread and the cluster cases, providing small volumes of uncertainty for low number of windows and moderate directional dependence.\\
%%%%%%%%%%%%%%%%%%%%%%%%%%%%%%%%%%%%%%%%%%%%%%%%%%%%%%%%%%%%%%%%%%%%
\begin{figure}
    \centering
    %\fbox{\rule{10cm}{0pt}\rule[-0.5ex]{0pt}{8cm}} %placeholder
    \includegraphics{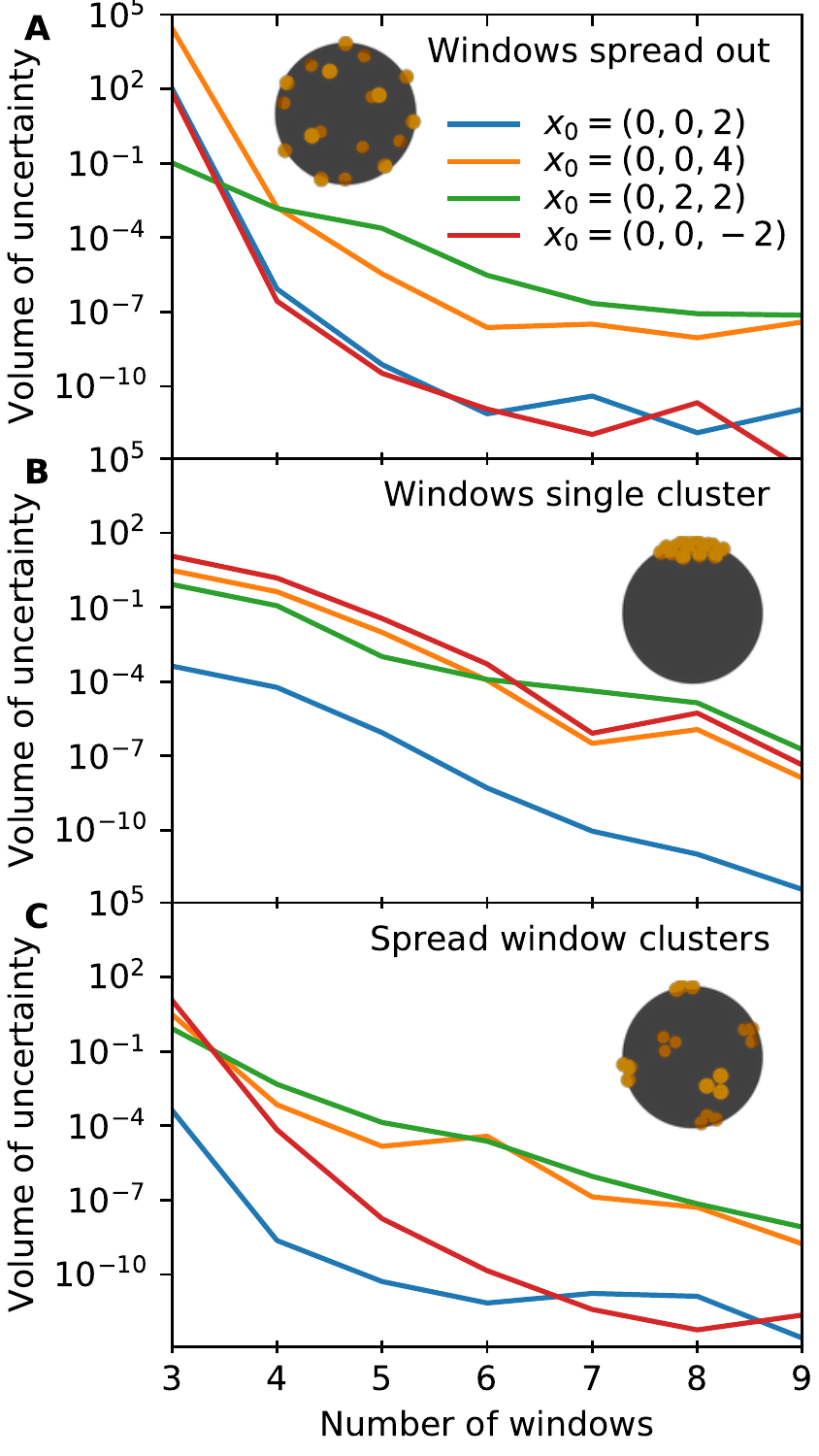}
    \caption{{\bf Uncertainty of the source location} Source location recovery is subject to measurement error in the fluxes which is characterized by a volume of uncertainty $V_u$, depending on the number $M$ and location of the windows as well as the position of the source. {\bf (A)} $V_u$ is given by the volume of intersection of the error parallelepipeds of all combinations of three windows and thus decreases with the number of windows. {\bf (B)} Windows spread uniformly across the ball surface yields high $V_u$ for low $M$ but low directional dependence overall. {\bf (B)} A single cluster of windows yields very low $V_u$ at low $M$ but a high directional dependence. {\bf (C)} Window clusters spread uniformly across the ball surface (at most three windows per cluster) provides a compromise for lower $V_u$ at low $M$ and low directional dependence. Overall, the volume of uncertainty curves span a surprisingly wide range.}
    \label{fig:figure4}
\end{figure}
%%%%%%%%%%%%%%%%%%%%%%%%%%%%%%%%%%%%%%%%%%%%%%%%%%%%%%%%%%%%%%%%%%%%
{\noindent \bf Concluding remarks}
Diffusing cues arriving at narrow windows located on the surface of ball is a model of cell sensing. We show here that at least three windows are necessary to triangulate the position of the source from the diffusion steady-state fluxes. Interestingly, fluxes at additional receptors increases drastically the precision of the source. However, in three dimensions, the source cannot be detected too far from the cell compared to two dimensions or when the cell sensing region is placed in a narrow tube. Possibly recovering a source further away could be possible when binding is not infinitely fast. \\
The present framework  to recover the source of a gradient is general and could be relevant to any cells, bacteria, growth cones \cite{chedotal2010wiring} that have to find a target. This first step should be followed by a second step of transduction that preserve the quantitative difference of signal measured at each receptors \cite{bouzigues2007asymmetric,bouzigues2010mechanism}. Another possible improvement of the model would be to consider multiple synergetic cues \cite{shukron2019chemical}.
%%%%%%%%%%%%%%%%%%%%%%%%%%%%%%%%%%%%%%%%%%%%%%%%%%%%%%%%%%%%%%%%%%%%
\bibliography{PRLCellsensingbiblio2,RMPbiblio4newN2}% Produces the bibliography via BibTeX.

%merlin.mbs apsrev4-1.bst 2010-07-25 4.21a (PWD, AO, DPC) hacked
%Control: key (0)
%Control: author (8) initials jnrlst
%Control: editor formatted (1) identically to author
%Control: production of article title (-1) disabled
%Control: page (0) single
%Control: year (1) truncated
%Control: production of eprint (0) enabled
\begin{thebibliography}{22}%
\makeatletter
\providecommand \@ifxundefined [1]{%
 \@ifx{#1\undefined}
}%
\providecommand \@ifnum [1]{%
 \ifnum #1\expandafter \@firstoftwo
 \else \expandafter \@secondoftwo
 \fi
}%
\providecommand \@ifx [1]{%
 \ifx #1\expandafter \@firstoftwo
 \else \expandafter \@secondoftwo
 \fi
}%
\providecommand \natexlab [1]{#1}%
\providecommand \enquote  [1]{``#1''}%
\providecommand \bibnamefont  [1]{#1}%
\providecommand \bibfnamefont [1]{#1}%
\providecommand \citenamefont [1]{#1}%
\providecommand \href@noop [0]{\@secondoftwo}%
\providecommand \href [0]{\begingroup \@sanitize@url \@href}%
\providecommand \@href[1]{\@@startlink{#1}\@@href}%
\providecommand \@@href[1]{\endgroup#1\@@endlink}%
\providecommand \@sanitize@url [0]{\catcode `\\12\catcode `\$12\catcode
  `\&12\catcode `\#12\catcode `\^12\catcode `\_12\catcode `\%12\relax}%
\providecommand \@@startlink[1]{}%
\providecommand \@@endlink[0]{}%
\providecommand \url  [0]{\begingroup\@sanitize@url \@url }%
\providecommand \@url [1]{\endgroup\@href {#1}{\urlprefix }}%
\providecommand \urlprefix  [0]{URL }%
\providecommand \Eprint [0]{\href }%
\providecommand \doibase [0]{http://dx.doi.org/}%
\providecommand \selectlanguage [0]{\@gobble}%
\providecommand \bibinfo  [0]{\@secondoftwo}%
\providecommand \bibfield  [0]{\@secondoftwo}%
\providecommand \translation [1]{[#1]}%
\providecommand \BibitemOpen [0]{}%
\providecommand \bibitemStop [0]{}%
\providecommand \bibitemNoStop [0]{.\EOS\space}%
\providecommand \EOS [0]{\spacefactor3000\relax}%
\providecommand \BibitemShut  [1]{\csname bibitem#1\endcsname}%
\let\auto@bib@innerbib\@empty
%</preamble>
\bibitem [{\citenamefont {Kaupp}\ and\ \citenamefont
  {Str{\"u}nker}(2017)}]{kaupp2017signaling}%
  \BibitemOpen
  \bibfield  {author} {\bibinfo {author} {\bibfnamefont {U.~B.}\ \bibnamefont
  {Kaupp}}\ and\ \bibinfo {author} {\bibfnamefont {T.}~\bibnamefont
  {Str{\"u}nker}},\ }\href@noop {} {\bibfield  {journal} {\bibinfo  {journal}
  {Trends in cell biology}\ }\textbf {\bibinfo {volume} {27}},\ \bibinfo
  {pages} {101} (\bibinfo {year} {2017})}\BibitemShut {NoStop}%
\bibitem [{\citenamefont {Str{\"u}nker}\ \emph {et~al.}(2015)\citenamefont
  {Str{\"u}nker}, \citenamefont {Alvarez},\ and\ \citenamefont
  {Kaupp}}]{strunker2015physical}%
  \BibitemOpen
  \bibfield  {author} {\bibinfo {author} {\bibfnamefont {T.}~\bibnamefont
  {Str{\"u}nker}}, \bibinfo {author} {\bibfnamefont {L.}~\bibnamefont
  {Alvarez}}, \ and\ \bibinfo {author} {\bibfnamefont {U.}~\bibnamefont
  {Kaupp}},\ }\href@noop {} {\bibfield  {journal} {\bibinfo  {journal} {Current
  opinion in neurobiology}\ }\textbf {\bibinfo {volume} {34}},\ \bibinfo
  {pages} {110} (\bibinfo {year} {2015})}\BibitemShut {NoStop}%
\bibitem [{\citenamefont {Wachten}\ \emph {et~al.}(2017)\citenamefont
  {Wachten}, \citenamefont {Jikeli},\ and\ \citenamefont
  {Kaupp}}]{wachten2017sperm}%
  \BibitemOpen
  \bibfield  {author} {\bibinfo {author} {\bibfnamefont {D.}~\bibnamefont
  {Wachten}}, \bibinfo {author} {\bibfnamefont {J.~F.}\ \bibnamefont {Jikeli}},
  \ and\ \bibinfo {author} {\bibfnamefont {U.~B.}\ \bibnamefont {Kaupp}},\
  }\href@noop {} {\bibfield  {journal} {\bibinfo  {journal} {Cold Spring Harbor
  perspectives in biology}\ }\textbf {\bibinfo {volume} {9}},\ \bibinfo {pages}
  {a028225} (\bibinfo {year} {2017})}\BibitemShut {NoStop}%
\bibitem [{\citenamefont {Alvarez}\ \emph {et~al.}(2014)\citenamefont
  {Alvarez}, \citenamefont {Friedrich}, \citenamefont {Gompper},\ and\
  \citenamefont {Kaupp}}]{alvarez2014computational}%
  \BibitemOpen
  \bibfield  {author} {\bibinfo {author} {\bibfnamefont {L.}~\bibnamefont
  {Alvarez}}, \bibinfo {author} {\bibfnamefont {B.~M.}\ \bibnamefont
  {Friedrich}}, \bibinfo {author} {\bibfnamefont {G.}~\bibnamefont {Gompper}},
  \ and\ \bibinfo {author} {\bibfnamefont {U.~B.}\ \bibnamefont {Kaupp}},\
  }\href@noop {} {\bibfield  {journal} {\bibinfo  {journal} {Trends in cell
  biology}\ }\textbf {\bibinfo {volume} {24}},\ \bibinfo {pages} {198}
  (\bibinfo {year} {2014})}\BibitemShut {NoStop}%
\bibitem [{\citenamefont {Ch{\'e}dotal}\ and\ \citenamefont
  {Richards}(2010)}]{chedotal2010wiring}%
  \BibitemOpen
  \bibfield  {author} {\bibinfo {author} {\bibfnamefont {A.}~\bibnamefont
  {Ch{\'e}dotal}}\ and\ \bibinfo {author} {\bibfnamefont {L.~J.}\ \bibnamefont
  {Richards}},\ }\href@noop {} {\bibfield  {journal} {\bibinfo  {journal} {Cold
  Spring Harbor perspectives in biology}\ ,\ \bibinfo {pages} {a001917}}
  (\bibinfo {year} {2010})}\BibitemShut {NoStop}%
\bibitem [{\citenamefont {Goodhill}(2016)}]{goodhill2016can}%
  \BibitemOpen
  \bibfield  {author} {\bibinfo {author} {\bibfnamefont {G.~J.}\ \bibnamefont
  {Goodhill}},\ }\href@noop {} {\bibfield  {journal} {\bibinfo  {journal}
  {Trends in neurosciences}\ }\textbf {\bibinfo {volume} {39}},\ \bibinfo
  {pages} {202} (\bibinfo {year} {2016})}\BibitemShut {NoStop}%
\bibitem [{\citenamefont {Reingruber}\ and\ \citenamefont
  {Holcman}(2014)}]{reingruber2014computational}%
  \BibitemOpen
  \bibfield  {author} {\bibinfo {author} {\bibfnamefont {J.}~\bibnamefont
  {Reingruber}}\ and\ \bibinfo {author} {\bibfnamefont {D.}~\bibnamefont
  {Holcman}},\ }in\ \href@noop {} {\emph {\bibinfo {booktitle} {Seminars in
  cell \& developmental biology}}},\ Vol.~\bibinfo {volume} {35}\ (\bibinfo
  {organization} {Elsevier},\ \bibinfo {year} {2014})\ pp.\ \bibinfo {pages}
  {189--202}\BibitemShut {NoStop}%
\bibitem [{\citenamefont {Gorelashvili}\ \emph {et~al.}(2014)\citenamefont
  {Gorelashvili}, \citenamefont {Emmert}, \citenamefont {Hodeck},\ and\
  \citenamefont {Heinrich}}]{gorelashvili2014amoeboid}%
  \BibitemOpen
  \bibfield  {author} {\bibinfo {author} {\bibfnamefont {M.}~\bibnamefont
  {Gorelashvili}}, \bibinfo {author} {\bibfnamefont {M.}~\bibnamefont
  {Emmert}}, \bibinfo {author} {\bibfnamefont {K.~F.}\ \bibnamefont {Hodeck}},
  \ and\ \bibinfo {author} {\bibfnamefont {D.}~\bibnamefont {Heinrich}},\
  }\href@noop {} {\bibfield  {journal} {\bibinfo  {journal} {New journal of
  physics}\ }\textbf {\bibinfo {volume} {16}},\ \bibinfo {pages} {075012}
  (\bibinfo {year} {2014})}\BibitemShut {NoStop}%
\bibitem [{\citenamefont {Wondergem}\ \emph {et~al.}(2019)\citenamefont
  {Wondergem}, \citenamefont {Mytiliniou}, \citenamefont {de~Wit},
  \citenamefont {Reuvers}, \citenamefont {Holcman},\ and\ \citenamefont
  {Heinrich}}]{wondergem2019chemotaxis}%
  \BibitemOpen
  \bibfield  {author} {\bibinfo {author} {\bibfnamefont {J.~A.}\ \bibnamefont
  {Wondergem}}, \bibinfo {author} {\bibfnamefont {M.}~\bibnamefont
  {Mytiliniou}}, \bibinfo {author} {\bibfnamefont {F.~C.}\ \bibnamefont
  {de~Wit}}, \bibinfo {author} {\bibfnamefont {T.~G.}\ \bibnamefont {Reuvers}},
  \bibinfo {author} {\bibfnamefont {D.}~\bibnamefont {Holcman}}, \ and\
  \bibinfo {author} {\bibfnamefont {D.}~\bibnamefont {Heinrich}},\ }\href@noop
  {} {\bibfield  {journal} {\bibinfo  {journal} {bioRxiv}\ ,\ \bibinfo {pages}
  {735779}} (\bibinfo {year} {2019})}\BibitemShut {NoStop}%
\bibitem [{\citenamefont {Berg}\ and\ \citenamefont
  {Purcell}(1977)}]{Berg1977}%
  \BibitemOpen
  \bibfield  {author} {\bibinfo {author} {\bibfnamefont {H.~C.}\ \bibnamefont
  {Berg}}\ and\ \bibinfo {author} {\bibfnamefont {M.}~\bibnamefont {Purcell}},\
  }\href@noop {} {\bibfield  {journal} {\bibinfo  {journal} {Biophys. J.}\
  }\textbf {\bibinfo {volume} {20}},\ \bibinfo {pages} {193} (\bibinfo {year}
  {1977})}\BibitemShut {NoStop}%
\bibitem [{\citenamefont {Zwanzig}(1990)}]{Zwanzig}%
  \BibitemOpen
  \bibfield  {author} {\bibinfo {author} {\bibfnamefont {R.}~\bibnamefont
  {Zwanzig}},\ }\href {\doibase 10.1073/pnas.87.15.5856} {\bibfield  {journal}
  {\bibinfo  {journal} {Proceedings of the National Academy of Sciences of the
  United States of America}\ }\textbf {\bibinfo {volume} {87}},\ \bibinfo
  {pages} {5856} (\bibinfo {year} {1990})}\BibitemShut {NoStop}%
\bibitem [{\citenamefont {Bialek}\ and\ \citenamefont
  {Setayeshgar}(2008)}]{bialek2008cooperativity}%
  \BibitemOpen
  \bibfield  {author} {\bibinfo {author} {\bibfnamefont {W.}~\bibnamefont
  {Bialek}}\ and\ \bibinfo {author} {\bibfnamefont {S.}~\bibnamefont
  {Setayeshgar}},\ }\href@noop {} {\bibfield  {journal} {\bibinfo  {journal}
  {Physical Review Letters}\ }\textbf {\bibinfo {volume} {100}},\ \bibinfo
  {pages} {258101} (\bibinfo {year} {2008})}\BibitemShut {NoStop}%
\bibitem [{\citenamefont {Endres}\ and\ \citenamefont
  {Wingreen}(2008)}]{endres2008accuracy}%
  \BibitemOpen
  \bibfield  {author} {\bibinfo {author} {\bibfnamefont {R.~G.}\ \bibnamefont
  {Endres}}\ and\ \bibinfo {author} {\bibfnamefont {N.~S.}\ \bibnamefont
  {Wingreen}},\ }\href@noop {} {\bibfield  {journal} {\bibinfo  {journal}
  {Proceedings of the National Academy of Sciences}\ }\textbf {\bibinfo
  {volume} {105}},\ \bibinfo {pages} {15749} (\bibinfo {year}
  {2008})}\BibitemShut {NoStop}%
\bibitem [{\citenamefont {Kaizu}\ \emph {et~al.}(2014)\citenamefont {Kaizu},
  \citenamefont {De~Ronde}, \citenamefont {Paijmans}, \citenamefont
  {Takahashi}, \citenamefont {Tostevin},\ and\ \citenamefont
  {Ten~Wolde}}]{kaizu2014berg}%
  \BibitemOpen
  \bibfield  {author} {\bibinfo {author} {\bibfnamefont {K.}~\bibnamefont
  {Kaizu}}, \bibinfo {author} {\bibfnamefont {W.}~\bibnamefont {De~Ronde}},
  \bibinfo {author} {\bibfnamefont {J.}~\bibnamefont {Paijmans}}, \bibinfo
  {author} {\bibfnamefont {K.}~\bibnamefont {Takahashi}}, \bibinfo {author}
  {\bibfnamefont {F.}~\bibnamefont {Tostevin}}, \ and\ \bibinfo {author}
  {\bibfnamefont {P.~R.}\ \bibnamefont {Ten~Wolde}},\ }\href@noop {} {\bibfield
   {journal} {\bibinfo  {journal} {Biophysical journal}\ }\textbf {\bibinfo
  {volume} {106}},\ \bibinfo {pages} {976} (\bibinfo {year}
  {2014})}\BibitemShut {NoStop}%
\bibitem [{\citenamefont {Aquino}\ \emph {et~al.}(2016)\citenamefont {Aquino},
  \citenamefont {Wingreen},\ and\ \citenamefont {Endres}}]{aquino2016know}%
  \BibitemOpen
  \bibfield  {author} {\bibinfo {author} {\bibfnamefont {G.}~\bibnamefont
  {Aquino}}, \bibinfo {author} {\bibfnamefont {N.~S.}\ \bibnamefont
  {Wingreen}}, \ and\ \bibinfo {author} {\bibfnamefont {R.~G.}\ \bibnamefont
  {Endres}},\ }\href@noop {} {\bibfield  {journal} {\bibinfo  {journal}
  {Journal of statistical physics}\ }\textbf {\bibinfo {volume} {162}},\
  \bibinfo {pages} {1353} (\bibinfo {year} {2016})}\BibitemShut {NoStop}%
\bibitem [{\citenamefont {Dobramysl}\ and\ \citenamefont
  {Holcman}(2018{\natexlab{a}})}]{dobramysl2018mixed}%
  \BibitemOpen
  \bibfield  {author} {\bibinfo {author} {\bibfnamefont {U.}~\bibnamefont
  {Dobramysl}}\ and\ \bibinfo {author} {\bibfnamefont {D.}~\bibnamefont
  {Holcman}},\ }\href@noop {} {\bibfield  {journal} {\bibinfo  {journal}
  {Journal of computational physics}\ }\textbf {\bibinfo {volume} {355}},\
  \bibinfo {pages} {22} (\bibinfo {year} {2018}{\natexlab{a}})}\BibitemShut
  {NoStop}%
\bibitem [{\citenamefont {Dobramysl}\ and\ \citenamefont
  {Holcman}(2018{\natexlab{b}})}]{dobramysl2018reconstructing}%
  \BibitemOpen
  \bibfield  {author} {\bibinfo {author} {\bibfnamefont {U.}~\bibnamefont
  {Dobramysl}}\ and\ \bibinfo {author} {\bibfnamefont {D.}~\bibnamefont
  {Holcman}},\ }\href@noop {} {\bibfield  {journal} {\bibinfo  {journal}
  {Scientific reports}\ }\textbf {\bibinfo {volume} {8}},\ \bibinfo {pages}
  {941} (\bibinfo {year} {2018}{\natexlab{b}})}\BibitemShut {NoStop}%
\bibitem [{\citenamefont {Shukron}\ \emph {et~al.}(2019)\citenamefont
  {Shukron}, \citenamefont {Dobramysl},\ and\ \citenamefont
  {Holcman}}]{shukron2019chemical}%
  \BibitemOpen
  \bibfield  {author} {\bibinfo {author} {\bibfnamefont {O.}~\bibnamefont
  {Shukron}}, \bibinfo {author} {\bibfnamefont {U.}~\bibnamefont {Dobramysl}},
  \ and\ \bibinfo {author} {\bibfnamefont {D.}~\bibnamefont {Holcman}},\
  }\href@noop {} {\bibfield  {journal} {\bibinfo  {journal} {Chemical Kinetics:
  Beyond The Textbook}\ ,\ \bibinfo {pages} {353}} (\bibinfo {year}
  {2019})}\BibitemShut {NoStop}%
\bibitem [{\citenamefont {Lagache}\ and\ \citenamefont
  {Holcman}(2017)}]{lagache2017extended}%
  \BibitemOpen
  \bibfield  {author} {\bibinfo {author} {\bibfnamefont {T.}~\bibnamefont
  {Lagache}}\ and\ \bibinfo {author} {\bibfnamefont {D.}~\bibnamefont
  {Holcman}},\ }\href@noop {} {\bibfield  {journal} {\bibinfo  {journal}
  {Journal of Statistical Physics}\ }\textbf {\bibinfo {volume} {166}},\
  \bibinfo {pages} {244} (\bibinfo {year} {2017})}\BibitemShut {NoStop}%
\bibitem [{\citenamefont {Lindsay}\ \emph {et~al.}(2017)\citenamefont
  {Lindsay}, \citenamefont {Bernoff},\ and\ \citenamefont
  {Ward}}]{lindsay2017first}%
  \BibitemOpen
  \bibfield  {author} {\bibinfo {author} {\bibfnamefont {A.~E.}\ \bibnamefont
  {Lindsay}}, \bibinfo {author} {\bibfnamefont {A.~J.}\ \bibnamefont
  {Bernoff}}, \ and\ \bibinfo {author} {\bibfnamefont {M.~J.}\ \bibnamefont
  {Ward}},\ }\href@noop {} {\bibfield  {journal} {\bibinfo  {journal}
  {Multiscale Modeling \& Simulation}\ }\textbf {\bibinfo {volume} {15}},\
  \bibinfo {pages} {74} (\bibinfo {year} {2017})}\BibitemShut {NoStop}%
\bibitem [{\citenamefont {Bouzigues}\ \emph {et~al.}(2007)\citenamefont
  {Bouzigues}, \citenamefont {Morel}, \citenamefont {Triller},\ and\
  \citenamefont {Dahan}}]{bouzigues2007asymmetric}%
  \BibitemOpen
  \bibfield  {author} {\bibinfo {author} {\bibfnamefont {C.}~\bibnamefont
  {Bouzigues}}, \bibinfo {author} {\bibfnamefont {M.}~\bibnamefont {Morel}},
  \bibinfo {author} {\bibfnamefont {A.}~\bibnamefont {Triller}}, \ and\
  \bibinfo {author} {\bibfnamefont {M.}~\bibnamefont {Dahan}},\ }\href@noop {}
  {\bibfield  {journal} {\bibinfo  {journal} {Proceedings of the National
  Academy of Sciences}\ }\textbf {\bibinfo {volume} {104}},\ \bibinfo {pages}
  {11251} (\bibinfo {year} {2007})}\BibitemShut {NoStop}%
\bibitem [{\citenamefont {Bouzigues}\ \emph {et~al.}(2010)\citenamefont
  {Bouzigues}, \citenamefont {Holcman},\ and\ \citenamefont
  {Dahan}}]{bouzigues2010mechanism}%
  \BibitemOpen
  \bibfield  {author} {\bibinfo {author} {\bibfnamefont {C.}~\bibnamefont
  {Bouzigues}}, \bibinfo {author} {\bibfnamefont {D.}~\bibnamefont {Holcman}},
  \ and\ \bibinfo {author} {\bibfnamefont {M.}~\bibnamefont {Dahan}},\
  }\href@noop {} {\bibfield  {journal} {\bibinfo  {journal} {PLoS One}\
  }\textbf {\bibinfo {volume} {5}},\ \bibinfo {pages} {e9243} (\bibinfo {year}
  {2010})}\BibitemShut {NoStop}%
\end{thebibliography}%
\end{document}